\documentclass[11pt,a4paper]{article}

\newcommand{\hb} {\mbox{\sffamily HERA\,\protect\rule[.5ex]{1.ex}{.11ex}\,B}\ }
\newcommand{\hbp}{\mbox{\sffamily HERA\,\protect\rule[.5ex]{1.ex}{.11ex}\,B}}
\newcommand{\PJgy}{\mbox{\ensuremath{\mathrm{J}\mskip -2mu/\mskip -2mu\psi}}}
\newcommand{\jpsi}{\PJgy{}}
\newcommand{\rmv}{RICH Multiplicity Veto\ }
\newcommand{\rmvp}{RICH Multiplicity Veto}
\usepackage{epsfig}
\usepackage{afterpage}

\begin{document}
\title{\Huge\bf A \rmv\\ for the HERA-B Experiment}
\author{U.~Husemann,$^{1,}$\thanks{\texttt{husemann@hep.physik.uni-siegen.de}}~
  M.~Adams,$^2$ 
  M.~B\"ocker,$^1$
  M.~Br\"uggemann,$^1$\\
  P.~Buchholz,$^1$  
  C.~Cruse,$^3$ 	
  Y.~Kolotaev,$^1$
}

\date{\normalsize$^1$\emph{Fachbereich Physik, Universit\"at Siegen, 
    Walter-Flex-Stra{\ss}e 3, D--57068 Siegen, Germany}\\
  $^2$\emph{Institut f\"ur Physik, Universit\"at Dortmund, 
    D--44221 Dortmund, Germany, \\now at: Aareal Bank AG, Wiesbaden, Germany}\\
  $^3$\emph{Institut f\"ur Physik, Universit\"at Dortmund, 
    D--44221 Dortmund, Germany, \\now at: Bundesamt f\"ur Wehrtechnik und
    Beschaffung, Koblenz, Germany}
}

\maketitle

\begin{flushright}
\vspace{-90mm}
\begin{minipage}[r]{0.15\textwidth}
\flushright\bf SI-2003-1
\end{minipage}
\vspace{80mm}

\end{flushright}

\begin{abstract}
We present the design and commissioning of a new multiplicity veto for the
\hb detector, a fixed-target spectrometer originally designed to study
the physics of $\mathrm{B}$ mesons in proton-nucleus interactions. 
The \hb trigger
is a message-driven multi-level track trigger. The first level trigger (FLT)
consists of custom-made electronics, and the higher trigger levels are
implemented as PC farms.  

The multiplicity veto has been designed to reject high-multiplicity 
events before they enter the trigger chain. A veto signal is generated
based on the comparison of the number of photons in part of the \hb 
ring-imaging \v Cerenkov counter (RICH) with a programmable threshold.

The \rmv is a modular system. First the hits in 256 detector
channels are summed by Base Sum Cards (BSC), then FED Sum Cards (FSC) sum the
subtotals of up to eight BSCs. Finally
 the Veto Board (VB) takes the veto decision
based on the sum of up to 14 FSCs.

The \rmv has been successfully installed and commissioned in
\hbp. The measured veto efficiency is $(99.9991\pm 0.0001)\%$, 
and the system is used
in the routine data-taking of the \hb experiment.
\end{abstract}
\begin{center}

Poster presented at the IEEE Nuclear Science Symposium 2002,\\ 
\vspace{-0.5mm}

Norfolk, Virginia, November 12--14, 2002
\vspace{5mm}

{\footnotesize This work was supported by the German 
  Bundesministerium f\"ur 
\vspace{-0.7mm}

Bildung und Forschung (BMBF) under the contract number 5HB1PEA/7.}
\end{center}

\section{Introduction}
\subsection{The \hb Experiment}
The \hb experiment at the electron-proton collider HERA at DESY in
Hamburg, Germany, is a fixed-target spectrometer with large angular acceptance.
 \hb has been designed to study neutral $\mathrm{B}$ mesons produced 
in proton-nucleus interactions. 
\hbp's current physics program aims at the measurement of the production of
$\mathrm{c\overline{c}}$ and $\mathrm{b\overline{b}}$ bound states 
in different target materials.
The \hb target consists of eight wires in the halo of the HERA proton beam.
The wires can be independently moved transverse to the beam to adjust 
the average number of proton-nucleus interactions per bunch crossing.

\begin{figure}[t]
\centering
\includegraphics[width=0.75\textwidth]{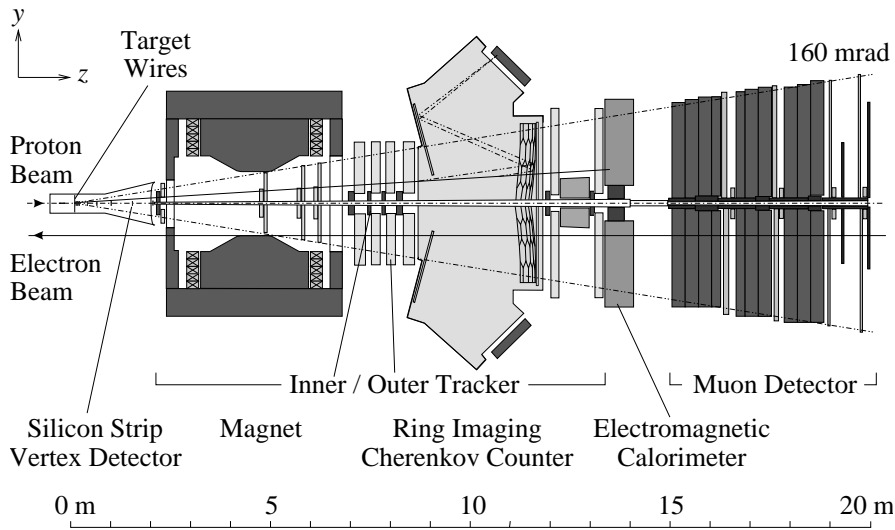}
\caption{\emph{Side view of the \hb detector}}
\label{fig:detector}
\end{figure}

The \hb detector is shown in Fig.~\ref{fig:detector}. It  
consists of tracking devices and devices used for particle
identification. The tracking devices comprise a vertex detector of
double-sided silicon micro-strip detectors, and an inner and outer tracking
system using micro-strip gaseous chambers with gas electron multiplier foils
(GEM-MSGCs) in the inner acceptance region ($10-100\,\mathrm{mrad}$ from the
proton beam direction) and
honey-comb drift chambers in the outer acceptance region
($100-220\,\mathrm{mrad}$). Particle identification is performed in the
electromagnetic calorimeter (ECAL), 
the ring-imaging \v Cerenkov
counter (RICH)~\cite{Arinho:2003} and the muon detector built 
of multi-wire proportional chambers.

To achieve the large suppression of background events needed 
for $\mathrm{c\overline{c}}$ and $\mathrm{b\overline{b}}$ 
production measurements, 
a highly selective multi-level trigger system is used. 
The first level trigger (FLT), built from custom-made
electronics, performs a track search
in the tracking detectors behind the spectrometer magnet. 
It has to sustain interaction rates of up to $20\,\mathrm{MHz}$. Track 
candidates for the FLT are provided by pretriggers in the muon 
detector and the ECAL, based on hit coincidences in adjacent detector layers
and hit clusters in the ECAL.
The \hb detector and trigger are described in greater
detail in~\cite{Abt:2002rd} and references therein.

\subsection{Motivation for a \rmv}
Large hit multiplicities are mainly caused by a superposition of 
multiple interactions in a single bunch crossing of the HERA proton beam.
The number of simultaneous interactions follows a Poisson distribution.

\begin{figure}[t]
\centering
\includegraphics[width=0.5\textwidth]{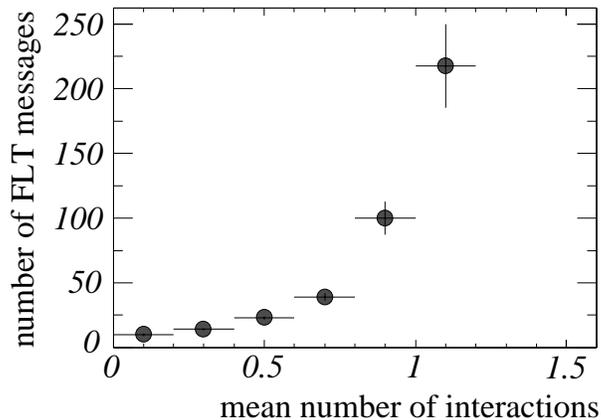}
\caption{\emph{Trigger rate of the FLT, measured by the number of FLT messages,
  as a function of the mean number of interactions per 
  bunch~\cite{Brueggemann:2002}. The number of
  FLT messages increases more than linearly with the number of interactions.}}
\label{fig:fltrate}
\end{figure}

The FLT is based on a track search algorithm implemented in custom-made
hardware which requires hits in four superlayers of the \hb tracking system.
Large hit multiplicities lead to an increase of the number of possible hit
combinations to form a track in the FLT. As a result, the FLT is
sensitive to events with large hit multiplicities, see Fig.~\ref{fig:fltrate}.
Since the FLT is a
modular message-driven system, the message load in its network
increases with the hit multiplicity which may introduce dead-time in the 
system. Therefore a mechanism
to protect the FLT from processing multi-interaction events is needed.
At the same time, this mechanism should preserve a large efficiency for
interesting physics signals, e.g. the decay $\jpsi \rightarrow \ell^+ \ell^-$.

\begin{figure}[t]
  \centering
  \includegraphics[width=0.75\textwidth]{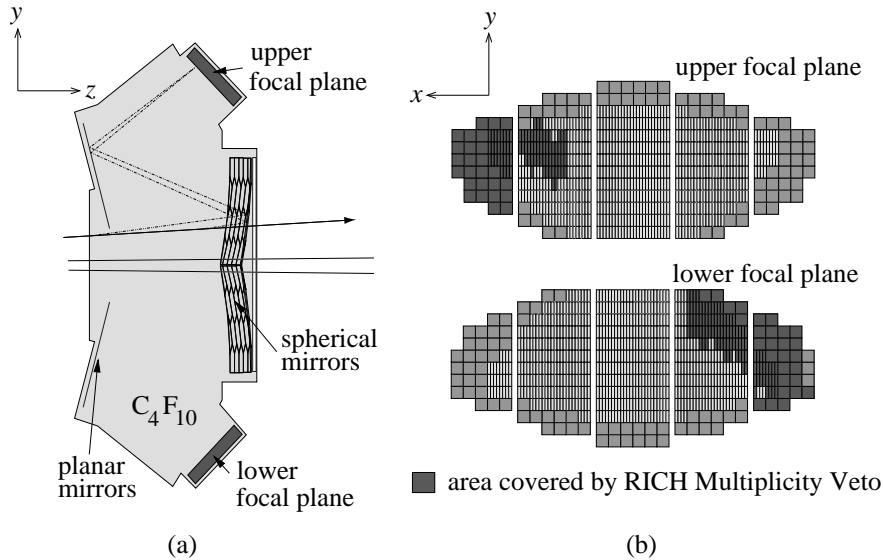}
  \caption{\emph{(a) The \hb RICH detector: Charged particles radiate \v Cerenkov
  photons in $\mathrm{C_4F_{10}}$ gas. Photons emitted in parallel cones 
  around the flight direction of a particle
  are reflected to the upper or lower focal plane by spherical and planar 
  mirrors, where they form a ring. (b) Focal planes of the RICH detector:
  The photons are detected by multi-anode photomultipliers of two sizes.
  The dark areas  are currently covered by the \rmv system.}}
  \label{fig:veto_area}
\end{figure}

\begin{figure}[t]
\centering
\includegraphics[width=0.75\textwidth]{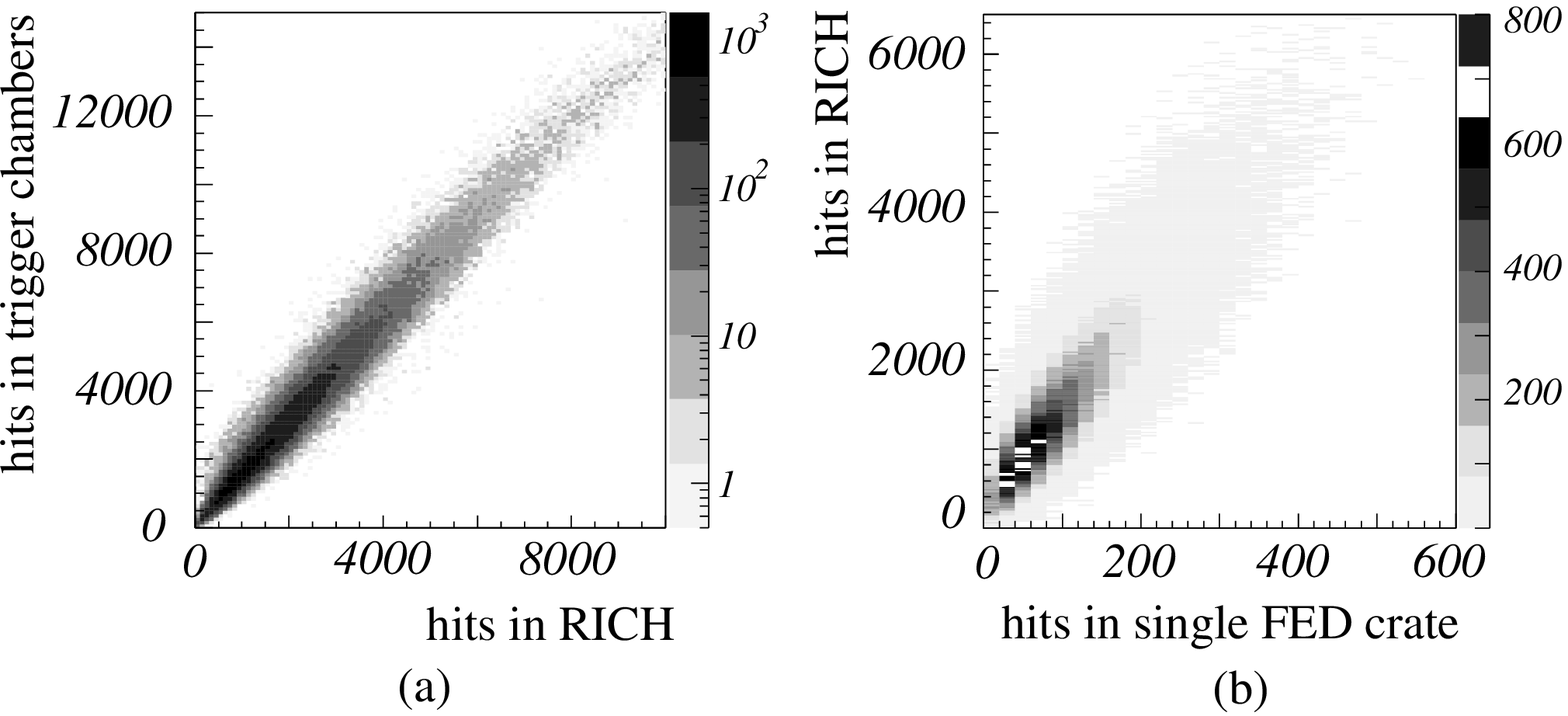}
\caption{\emph{(a) Correlation of the number of photons in part of the RICH 
  detector with the number of hits in the tracking chambers used in the
  first level trigger (FLT)~\cite{Adams:2001}.
  (b) Correlation of the number of photons processed by a single
  Front End Driver (FED) crate in the RICH detector with the number of
  photons in the whole RICH detector, corresponding to 14 FED 
  crates~\cite{Brueggemann:2002}.}}
\label{fig:correlation}
\end{figure}

A suitable measure of the hit multiplicity in the \hb detector is the 
number of photons generated by charged particles in the RICH. A sketch
of the \hb RICH detector can be found in Fig.~\ref{fig:veto_area}\,(a).
As shown in 
Fig.~\ref{fig:correlation}\,(a), the number
of RICH photons is highly correlated with the number of hits in the tracking 
chambers used in the FLT.
Even in the case when only a small part of the RICH detector is covered,
the number of RICH photons in this part is highly correlated with the number of
hits in the whole RICH, see Fig.~\ref{fig:correlation}\,(b).
Hence the veto functionality can already be established with partial
coverage of the RICH.

The RICH detector is suited to host a multiplicity veto not only from the
physics but also from the technological point of view. It has proven 
to be a very
reliable, low-noise system. In addition, its interface to the \hb data
acquisition system (DAQ), the front-end driver (FED), is easily accessible
to install the necessary additional electronics.

\subsection{Veto Strategy}
The strategy of a multiplicity veto using the RICH detector is to provide a
fast sum of the number of photons in part of the RICH and to generate a veto
signal
based on the comparison of this number of photons with an adjustable threshold.
In order to stop the trigger chain even before the message-driven processing
in the FLT, the veto signal has to be distributed to the pretriggers.

\subsection{Design Concepts}
\enlargethispage{3mm}
The photon signals detected by the photomultipliers of the RICH detector are
provided as digitized data at the FEDs. Therefore the summation of hits is
performed by digital electronics. The logic of the veto system is
implemented in large Complex Programmable Logic Devices 
(CPLD)~\cite{bib:alters} to 
allow for rapid development of the logic and to keep the design flexible.
The CPLDs used in the design are well-suited for cascaded summation.

Since the RICH FEDs are distributed in 14 FED crates close to the \hb 
detector, the veto system has to be a distributed
modular system. The current implementation of the \rmv covers two
FED crates, see Fig.~\ref{fig:veto_area}\,(b).

The maximum allowed latency $\tau_\mathrm{max}$ 
of the system is given by the conditional requirement that the pretrigger
systems have to be stopped before transfering data to the FLT.
Latency measurements show that the maximum allowed latency 
is given by the muon pretrigger. The veto signal has to arrive at the muon 
pretrigger within $\tau_\mathrm{max}=1067\,\mathrm{ns}$ 
after the interaction at the \hb target.

The \rmv is a new hardware component to be integrated into an
existing detector. Therefore care is taken not to introduce 
additional electronic noise into
the detector front-end. The electronics boards that receive
 the data from the FEDs are plugged directly on the FED boards 
in order to keep signal paths short. In addition, the signal lines 
on all electronics boards are routed carefully.

The data inside the multiplicity veto has to be transmitted over
approximately $40\,\mathrm{m}$ from the detector front-end to the
electronics trailer which is situated outside the experimental area.
Differential PECL logic is used for all flat cable connections to improve
the noise immunity.

Various specialized test boards and internal test facilities are needed to
allow for fast and controlled commissioning of the veto system.
\afterpage{\clearpage}

\section{The Electronics Boards}

\begin{figure}
\centering
\includegraphics[width=0.75\textwidth]{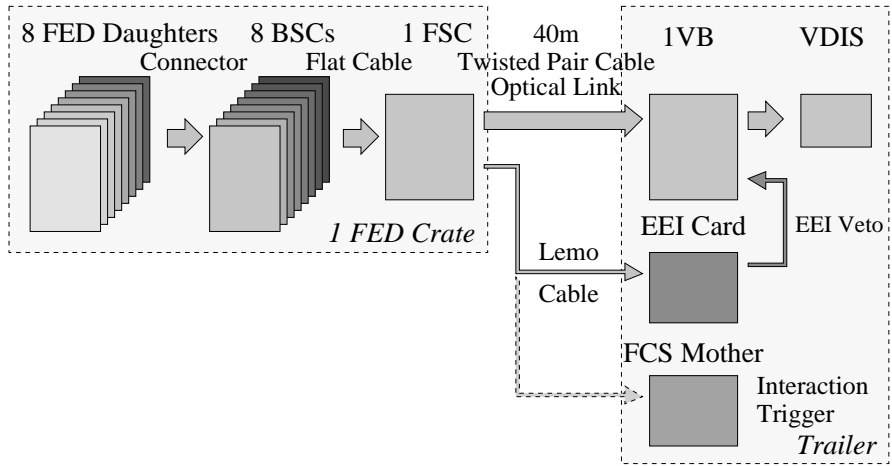}
\caption{\emph{Overview of the \rmv system. The components of the system are
  described in the text.}}
\label{fig:vetosystem}
\end{figure}

An overview of the \rmv system is shown in
Fig.~\ref{fig:vetosystem}. The RICH FED is distributed over 14 FED crates
with up to eight FED daughter cards processing data from 256 photomultiplier 
channels each. The data of eight FED daughter cards are transmitted to eight
Base Sum Cards (BSC), where they are summed. The sums of the eight
BSCs are transmitted to the FED Sum 
Card (FSC), which calculates the subtotal of the hits in 
a single RICH FED crate.
This sum is transmitted via twisted-pair flat cables or an optical link 
to the Veto Board (VB), which is situated in the \hb electronics trailer.
A modified I$^2$C bus system is used for communication among the modules
of the \rmv system.

The veto decision calculated from the sum of up to 14 FSC
subtotals is distributed to the pretrigger via the Veto Distributor 
(VDIS)~\cite{bib:vdis}.
A veto signal from the ECAL Energy Inhibit (EEI) board can be combined 
with the signal from the \rmv system.
The EEI is a veto system complementary to the \rmvp. It is based on the 
analog energy sum in the inner part of the ECAL. 

In addition, the veto signal generated from a single FSC can be fed into 
the Fast Control System (FCS) mother card, which distributes triggers
in the \hb experiment. In this ``Fast Veto'' mode, the veto system can 
be used as a low-bias interaction trigger. 

For the data-taking period 2002/2003, 15 BSCs, 2 FSCs, and the VB 
have been installed in the \hb detector.

\subsection{The Modified I$^2$C Bus}
A bus system for the \rmv has to address up to $14\times 9$ boards and to 
transmit signals over distances of more than $40\,\mathrm{m}$.
The Fast Control System~\cite{Fuljahn:1999}, 
the central bus system of the \hb experiment, cannot
be extended to fulfill these requirements. 

\begin{figure}[t]
  \centering
    \includegraphics[width=0.75\textwidth]{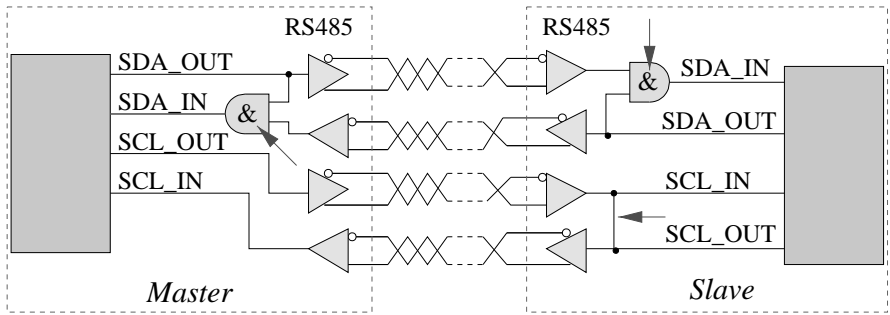}
  \caption{\emph{Modifications to the long-distance mode of the I$^2$C bus 
    controller, as indicated by the arrows: The I$^2$C bus controller needs
    a feedback of the data lines (SDA\_IN, SDA\_OUT) and the clock lines 
    (SCL\_IN, SCL\_OUT) to ensure the correct position of the acknowledge 
    signal after data transmission. The clock signals are only generated 
    by the master controller, hence SCL\_IN and SCL\_OUT can be connected
    on the slave controller side. The feedback on the signal lines is
    implemented as AND gates~\cite{Cruse:2002}.}}
  \label{fig:i2c_mod}
\end{figure}

The I$^2$C bus standard provides an easy and widely used protocol to 
implement the internal communication of the \rmv system. 
The original protocol is based on two bi-directional lines, one for
data transmission and one for the transmission of clock signals.
However, to use the I$^2$C bus in the \rmv
system, some modifications have been made to the protocol. The I$^2$C bus
controller used in the \rmvp~\cite{bib:i2c} is operated
in a special long-distance mode with four uni-directional lines. 
This mode allows to use long-distance drivers (RS485) for the I$^2$C lines.
In order to ensure the correct position of the acknowledge signal after
data transmission in the long-distance mode,
a feedback loop has to be introduced to 
the data and clock lines, as shown in Fig.~\ref{fig:i2c_mod}.

The I$^2$C master controller is situated on the VB. It is controlled via VME
commands, while the controllers on the BSCs and the FSCs are operated 
as I$^2$C slaves. Finite state machines implemented in dedicated CPLDs
on these boards are used to operate the I$^2$C slave controllers. 

The BSCs and the FSCs are entirely controlled via the I$^2$C bus. The bus
system is used to access  the command, status, and reference registers 
of the CPLDs on the BSCs and FSCs boards in order to initialize and monitor
these boards.

\subsection{The Base Sum Card}
The BSC is a six-layer electronics board which is plugged directly on an FED
daughter card. Almost the complete logic of the board is implemented in two
CPLDs. Each of the CPLDs processes signals from 128 of the 256 detector 
channels provided by the FED daughter card. 

\begin{figure}[t]
  \centering
  \includegraphics[width=0.75\textwidth]{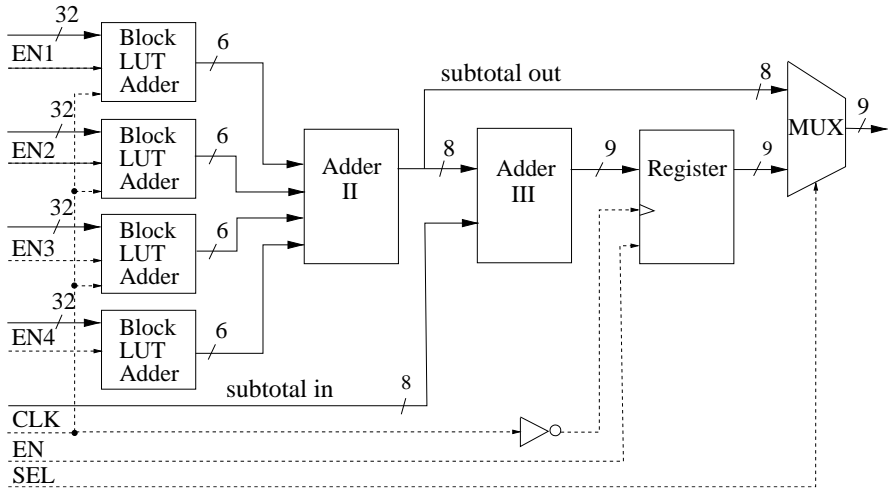}
  \caption{\emph{Logic of the summation CPLDs of the BSC~\cite{Cruse:2002}: 
    The subtotal of 32 FED
    channels is calculated in a block-LUT adder. Adder II sums the subtotals
    of four block-LUT adders. The result of this summation step can be 
    transmitted to the other CPLD via a multiplexer (MUX). The last adder unit
    (Adder III) of the other CPLD adds the result to its own subtotal, 
    before the final sum is
    written into the output register. Solid lines are data lines, 
    and dashed lines indicate steering signals.}}
\label{fig:summation}
\end{figure}

\begin{figure}[t]
\centering
\includegraphics[width=0.70\textwidth]{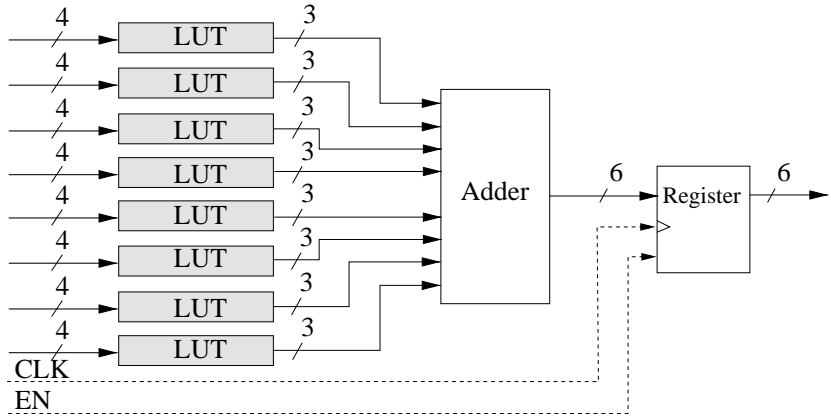}
\caption{\emph{Block-LUT adder~\cite{Cruse:2002}: 
  32 bits are divided in eight four-bit blocks. The
  number of set bits in a four-bit block 
  is determined via a look-up table (LUT),
  reducing the number of bits transferred to the adder unit to 24. 
  The adder performs a three-step cascaded sum, and the result,
  a six-bit number, is written to the output register.
  Solid lines are data lines, 
  and dashed lines indicate clock and steering signals.}}
\label{fig:blocklut}
\end{figure}

The summation logic of the CPLD is shown in Fig.~\ref{fig:summation}. The 128 
channels are further divided into four 32-bit blocks. 	
For every 32-bit block, a ``block-LUT adder'' unit is implemented, see 
Fig.~\ref{fig:blocklut}.
Eight look-up tables are used to count the number of hits in four bits and
transform them into a three-bit number. 
Each 32-bit block can be masked individually to account for problematic 
channels in the detector readout.

The subtotals of the four 32-bit blocks are summed by an additional adder
unit, and the subtotal of the second CPLD is added, before the sum is
transmitted to the FSC. 

In order to assign the veto signal generated by the \rmv to a given event, the
bunch crossing number (BX)
of the current event is transmitted to the FSC together
with the subtotal of the BSC. The BX is provided by the \hb FCS system and
transmitted to the BSC via the backplane of the FED crate.

While for the internal logic of the BSC TTL is used, the logic level
is transformed to differential PECL for data transmission. Due to a
limitation of number of lines on the flat cable, the least significant
bit (LSB) of the sum is discarded before the data transmission occurs.
\afterpage{\clearpage}

\subsection{The FED Sum Card}
\begin{figure}[t]
  \centering
  \includegraphics[width=0.52\textwidth]{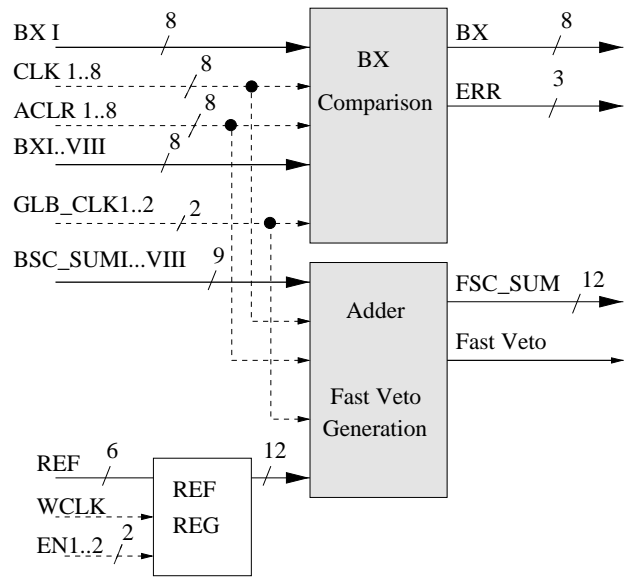}
  \caption{\emph{Summation logic of the FSC~\cite{Cruse:2002}: The subtotals of eight BSCs 
    (BSC\_SUMI..VIII) are summed by an adder unit. The resulting 12-bit sum
    can be compared with a reference value
    from the reference register (REFREG) to 
    generate a Fast Veto. The BX data of the eight BSCs (BXI..VIII) 
    are compared, and the
    BX of the first BSC is transmitted to the VB, together with a three-bit
    error code indicating the result of the BX comparison.
    Solid lines are data lines, and
    dashed lines indicate clock and steering signals.}}
  \label{fig:fsc}
\end{figure}

The FSC, situated in an FED crate together with up to eight BSCs, 
receives data from the BSCs via flat cables. The summation logic is depicted in 
Fig.~\ref{fig:fsc}. The subtotals of the BSCs are
added. To check the synchronization of the BSCs, the BXs which are transmitted
together with the subtotals are compared. In case of inconsistencies, an
error code is generated. The result of the summation and the error code
are transmitted to the VB.

The subtotal of the FSC is compared with an
adjustable threshold, and a ``Fast Veto'' signal is generated if the subtotal
is equal to or larger than the threshold. The Fast Veto signal can
be fed into the FCS mother card via a Lemo cable to be used as an 
interaction trigger, sensitive to minimal activity in part of the
RICH detector.

Since the data transmission from the FSC to the VB takes place
in a harsh hadronic environment, the FSC foresees data transmission via optical
links in addition to conventional twisted-pair flat cables.
In-situ tests show that using flat cables for the data transmission
in the \hb experimental area is feasible.
Moreover, in flat cable data transmission, no additional time
is needed to serialize the data for the optical transmission, thus
the latency of the system is smaller compared to optical transmission.
Therefore,  transmission via flat cables has been chosen for the 
installation of the \rmv system.

\subsection{The Veto Board}
The VB is a 6U electronics board designed in VME standard. 
It is situated in the \hb electronics trailer. 
The subtotals of up to 14 FSCs arrive at the input connectors
of the VB via flat cables. 
The VB is designed to consist of a main board and a mezzanine board,
each processing the data from seven FSCs. In the current installation 
in \hbp, only the main board is used.

\begin{figure}[t]
  \centering
  \includegraphics[width=0.75\textwidth]{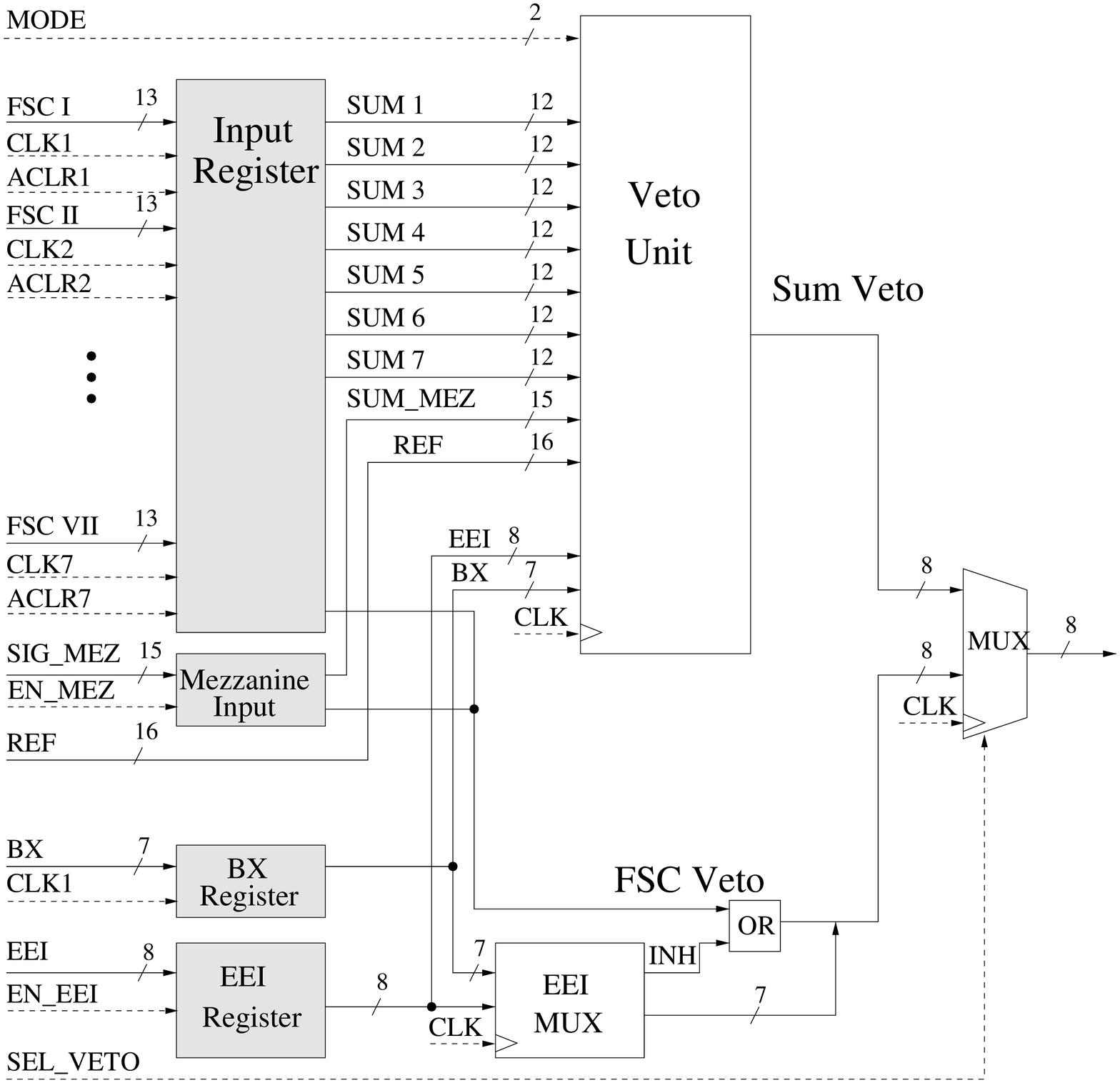}
  \caption{\emph{Veto logic of the VB (main board)~\cite{Cruse:2002}: 
    The subtotals of seven FSCs (FSC\,I..FSC\,VII) and the lower
    sum of the mezzanine board (SIG\_MEZ) are written to input registers.
    The veto signal from the EEI can be combined with the veto from the VB,
    and the BX of the first FSC is used to tag the veto signal with a
    BX. A multiplexer (MUX) allows to 
    choose between two veto sources: the veto unit, where the decision is
    based on the comparison of the total hit sum
    with two adjustable thresholds, or the logical
    OR of the Fast Veto signals from the FSCs. 
    Solid lines are signal lines, and dashed lines indicate clock and 
    steering signals.}}
  \label{fig:vetoflex}
\end{figure}

\begin{figure}[t]
  \centering
  \includegraphics[width=0.53\textwidth]{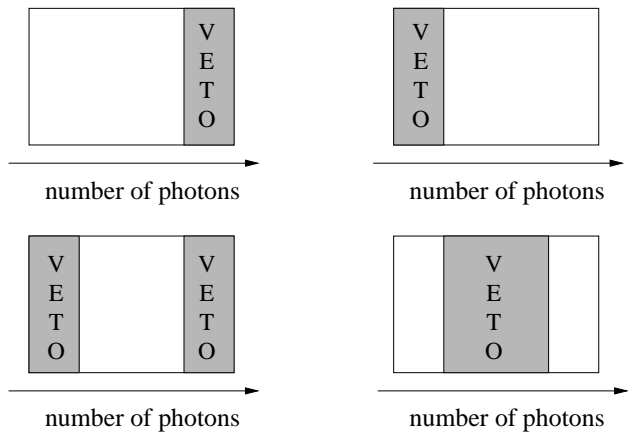}
  \caption{\emph{The VB allows for four different veto modi, 
    depending on the use of the upper and lower threshold.}}
  \label{fig:vetomodi}
\end{figure}

The veto logic of the VB is shown in Fig.~\ref{fig:vetoflex}.
The sums of seven FSCs and the final sum of the mezzanine board are
written to input registers before they are transmitted to the veto unit.
The veto unit calculates the final sum, and the veto signal is generated
based on a comparison with two adjustable thresholds. Hence, four different 
veto modi can be chosen, as depicted in Fig.~\ref{fig:vetomodi}. The veto
signal from the EEI can combined with the internal veto signal by a logical
OR.

The veto can also be generated from a different source: the logical OR of the
Fast Veto signals for the FSCs, also combined with the EEI veto,
can be used instead of the final sum.

The VME interface of the VB is used to control the entire \rmv
system, including the I$^2$C bus master controller and all
registers of the CPLDs of the VB and --
via the I$^2$C bus -- of the CPLDs of the BSCs and the FSCs.

The VB monitors the error codes generated by the FSCs due to BX
synchronization errors of the BSCs. In addition, the 
BX synchronization of the FSCs themselves and the number of vetos 
generated in the last 256 BX, both on the VB and on the EEI, are monitored.
This information is available in the status registers of a specialized
monitoring CPLD which can be read out via VME.

To transmit the veto decision and the veto rate to the \hb data
stream, the VB provides an interface to the Second Level Buffer (SLB),
\hbp's event buffer made of a network of specialized digital signal processors.
On the VB, a modified FED buffer is implemented. It contains the veto 
decision and the current veto mode, tagged with the FLT BX number, which is 
provided by the \hb FCS via the FCS backplane. 
If the FCS issues a trigger, the content of the FED buffer
is transmitted to the SLB and saved in the data stream. Due to this feature, 
the efficiency of the \rmv system can be determined easily, 
cf.\ Section~\ref{sec:efficiency}.

Since the internal monitoring provides the number of vetos only for the 
last 256 BX, i.e.\ on a statistical basis, a better monitoring mechanism is
needed to control the number of events rejected by the \rmv system. Thus the 
veto signal
is transmitted to a VME scaler which counts the number of vetos in an 
adjustable time interval. The scaler is read out via VME, and the measured
veto rate is written to the data stream.

\section{Commissioning}
The commissioning of the \rmv system comprises detailed
tests of the hardware functionality in the laboratory and measurements of
the latency of part of the system and the whole system. Furthermore, the
efficiency of the system is determined, and the influence on the \hb trigger
chain and DAQ and on physics signals is studied.

\subsection{Laboratory Tests}
\subsubsection{Bit Error Measurements}
The laboratory tests are based on a complex test setup using specialized
test boards to generate the input and analyze the output of the individual
boards so that the bit error level can be determined.

The BSC is tested with a specialized board that emulates the timing and
data of the \hb FED and the FCS, the Pretrigger FED Simulator (PFEDS).
 This board allows to generate defined test
input for the BSC. The test input is designed such that every channel is
tested individually, i.e.\ without bits being set in neighboring channels, 
hence hardware defects
can be located on the BSC. The output of the BSC is written to a FIFO on
the Base Sum Test Card (BSTC) where it can be read out via VME
and compared with the input. 

In the next step, the veto chain is extended by the FSC. With a special
adapter, the BSTC can also be used to test the output of the FSC. As for the 
BSC, every input channel is tested individually. By adding more BSCs,
the BX comparison and the summation on the FSC are tested. The
Fast Veto signal is tested by programming
 a threshold on the FSC via the I$^2$C 
bus and monitoring if the signal is generated correctly and assigned to the
correct BX. 

\begin{figure}[t]
\centering
\includegraphics[width=0.70\textwidth]{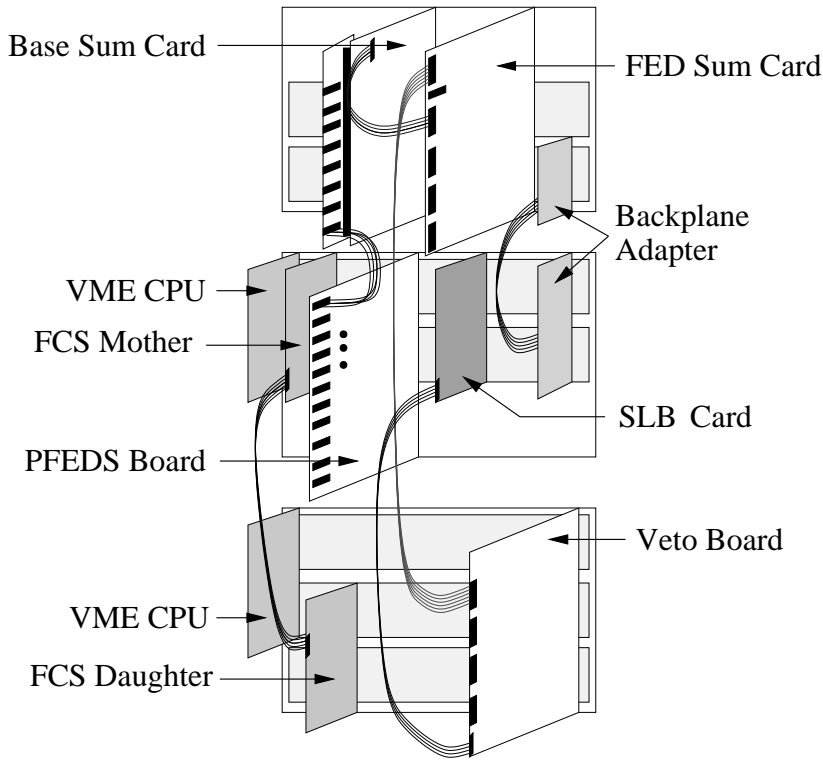}
\caption{\emph{Laboratory test setup to test the entire chain 
  of the \rmv system~\cite{Cruse:2002}. 
  A PFEDS board generates input data for the BSCs.
  Eight BSCs (only one shown in the figure) send their subtotals to an FSC, 
  and the FSC data is transmitted to the VB. 
  The VB is connected to an SLB card.
  The FCS mother generates BX and trigger signals and transmits them to the
  FCS daughter, which provides these signals for the VB. VME access to
  the VB, the FCS mother and the test modules is realized by two 
  VME CPUs.}}
\label{fig:systemtest}
\end{figure}

For tests of the VB, the Veto Board Test Card (VBTC) has been designed.
The VBTC uses PFEDS data as an input and emulates the output of two FSCs,
including the FSC subtotals, the Fast Veto signals, the BXs and the 
FSC error codes.
The timing of these signals can be shifted to emulate errors in the chain
of the veto system. An EEI emulator generates signals to test the insertion
of the EEI veto in the veto decision of the VB. 
For all four veto modi, a test of the correct generation of veto signals
has been performed.
The data transmission to the SLBs is tested with 
an SLB module, and the signal that triggers the data transmission
to the SLB is generated by the \hb FCS system. In Fig.~\ref{fig:systemtest}, 
a test setup for the entire chain of the RICH multiplicity veto system is 
shown.

The results of the laboratory 
tests are summarized in Table~\ref{tab:biterror}. 
During the tests, no bit errors could be detected, therefore
the error limits quoted in the table are given only by the test statistics.

\begin{table}[h]
\renewcommand{\arraystretch}{1.3}
\caption{\emph{Bit Error Measurements~\cite{Cruse:2002}}}
\label{tab:biterror}
\begin{center}
\begin{tabular}{cc}
\hline
Test Procedure & Bit Error Rate\\
\hline
BSC: summation &  $<1.6\times 10^{-8}$\\
FSC: summation&  $<1.1\times 10^{-10}$\\
FSC: BX comparison&  $<2.6\times 10^{-6}$\\
FSC: Fast Veto&  $<2.6\times 10^{-6}$\\
Veto Board & $<1.0\times 10^{-6}$\\
\hline
\end{tabular}
\end{center}
\end{table}

\subsubsection{Latency Measurements}
The system latency has been measured for each board individually.
The single board latency is defined as the time interval
between the rising edge of the input signal of the board 
and the rising edge of the output signal. 

The latency of the whole \rmv chain includes the time of flight 
from the target to the RICH detector, the latency of the FED system 
and additional latency introduced by the cable lengths.

The result of the latency measurement 
is summarized in Table~\ref{tab:latency}. The total latency 
$$\tau = (980\pm 2)\,\mathrm{ns}$$ 
is well 
below the maximum allowed value of $\tau_\mathrm{max} = 1067\,\mathrm{ns}$.

\begin{table}[h]
\renewcommand{\arraystretch}{1.3}
\caption{\emph{Latency Measurements~\cite{Cruse:2002}}}
\label{tab:latency}
\begin{center}
\begin{tabular}{cc}
  \hline
  Board & Latency [ns]\\
  \hline
  Base Sum Card &  $135\pm 2$\\
  FED Sum Card&  $110\pm 2$\\
  Veto Board & $173\pm 2$\\
  (Time of Flight, etc.) & $572\pm 2$\\
  \hline
  Total & $980 \pm 2 $\\
  \hline
\end{tabular}
\end{center}
\end{table}

\subsection{Efficiency Measurement}
\label{sec:efficiency}
The interface of the VB to the \hb data stream allows for an efficiency
measurement of the \rmv system~\cite{Brueggemann:2002}. The system can be
operated in ``spy mode'', i.e.\ the veto signal is not used to stop 
the pretriggers but the veto decision is saved in the data stream. 

The efficiency of the \rmv system is determined by comparing the veto 
decision made by the hardware with a simulation of the \rmv logic based 
on the recorded data. 
In the simulation, hits are counted from the recorded data in the
area of the RICH detector covered by the \rmv system.
After discarding the LSB of the subtotal for every installed BSC the
total hit sum is compared with the upper and lower
veto thresholds used in the data-taking.
This simulated veto decision is compared to the actual decision of the 
\rmv hardware. The efficiency $\varepsilon$ is defined as the fraction of 
correct veto decisions.

\begin{figure}[t]
\centering
\includegraphics[width=0.72\textwidth]{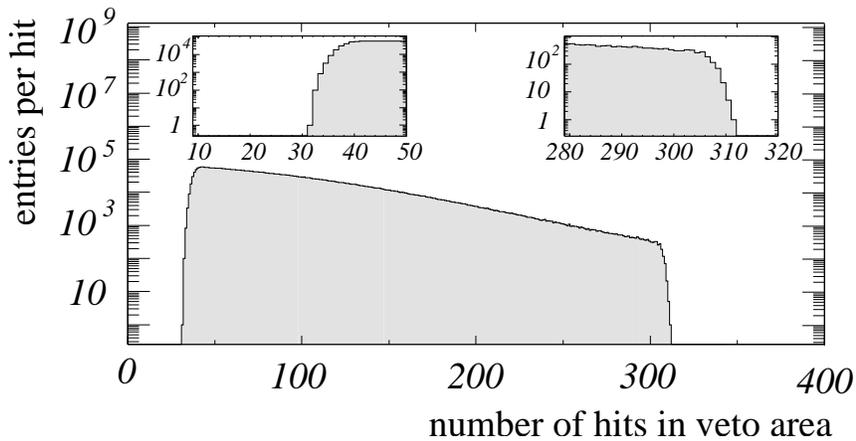}
\caption{\emph{Distribution of hits in the area covered by the \rmv for accepted 
  events~\cite{Brueggemann:2002}. 
  The \rmv was operated in a mode that accepts events inside a window
  of 30--300 hits. The hit distribution does not show 
  sharp edges, since the LSBs of 15 BSCs are discarded in the veto hardware.}}
\label{fig:efficiency}
\end{figure}

The efficiency measurement is based on a sample of 7.5 million events taken 
with a minimum bias trigger. Different veto thresholds
and veto modi have been used during the data-taking. 
As an example, in Fig.~\ref{fig:efficiency}, the hit distribution 
in the area covered by the \rmv system is shown for accepted events of a
run in which the veto accepted events inside a window of 30--300 hits. 

The  measurement shows that in very rare cases a wrong veto decision is
taken, but only if the number of hits is close to the threshold.
The efficiency averaged over all thresholds and modi is
$$\varepsilon = (99.9991 \pm 0.0001)\%.$$

\subsection{Influence on Trigger and DAQ}
\begin{figure}[t]
\centering
\includegraphics[width=0.50\textwidth]{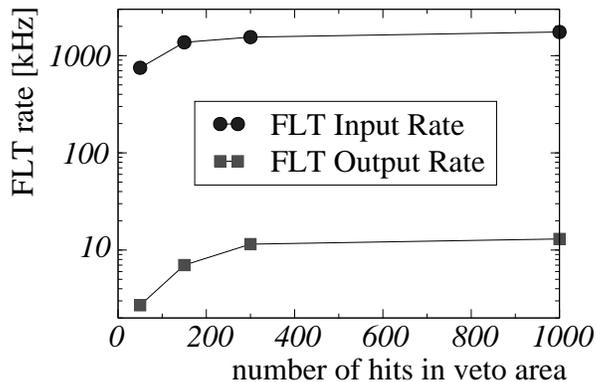}
\caption{\emph{FLT input and output rates as functions of the upper threshold of the
\rmvp.}}
\label{fig:fltrates}
\end{figure}

To test the influence of using the \rmv in the
data-taking, data is taken with the system actively rejecting events.
Some important 
parameters that determine the trigger and DAQ timing are measured as
functions of the upper veto threshold. As an example, the FLT input and
output rates are shown in Fig.~\ref{fig:fltrates}.  

The overall result is that all measured timing
parameters are improved. This includes smaller FLT input and output rates,
a smaller dead-time of the FCS, smaller event size 
and reconstruction time and hence a larger physics data rate to tape, cf.\ 
Table~\ref{tab:improve}.
These measurements show that using the \rmv protects the
trigger chain and the DAQ from high-multiplicity events.
Note however that the improvement values given in Table~\ref{tab:improve}
are only valid for a fixed set of trigger and DAQ settings,
which are subject to permanent optimization.

\begin{table}[h]
\renewcommand{\arraystretch}{1.3}
\caption{DAQ and Trigger Improvements (threshold: 300 hits)}
\label{tab:improve}
\begin{center}
\begin{tabular}{cc}
  \hline
  Value & Improvement\\
  \hline
  FLT input rate & 11\% \\
  FLT output rate & 12\% \\
  FCS dead-time & 50\% \\
  Event size & 10\%\\
  Reconstruction time & 35\%\\
  Data rate to tape & 25\%\\
  \hline
\end{tabular}
\end{center}
\end{table}

\subsection{Influence on Physics}
Using data taken during the \hb commissioning phase from 
August to October 2002, the influence of the \rmv on physics signals in the \hb
detector is studied.
The data sample includes approximately 3,500 candidates
for the decay $\jpsi\rightarrow\mu^+\mu^-$. The \rmv system has been inactive
during the data-taking, and cuts on the veto threshold are applied in the
offline analysis.
At a ``generic'' upper threshold
of 300 hits in the area covered by the \rmv system -- corresponding to
approximately 2,500 hits in the entire RICH -- the
survival of \jpsi{} mesons and a possible bias on the \jpsi{} kinematics are
studied.

\begin{figure}[t]
  \centering
  \includegraphics[width=0.50\textwidth]{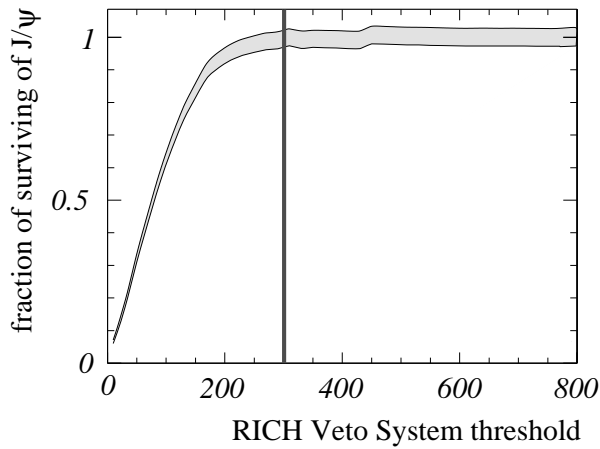}
  \caption{\emph{\jpsi{} survival fraction as a function of an upper
    threshold in the \rmv system. 
    The error band is given by the statistical error of
    the measurement of the number of \jpsi{} mesons. The vertical bar
    indicates the ``generic cut'' of 300 hits.}}
  \label{fig:njpsi}
\end{figure}

Fitting the di-muon invariant mass spectrum for events with less than a given
number of  hits and for all events, the survival fraction of \jpsi{} mesons 
is determined as a function of the RICH veto threshold, 
see Fig.~\ref{fig:njpsi}.
For an upper threshold of 300 hits, the survival fraction is
in the range of 97--100\%. 

\begin{figure}[t]
  \centering
  \includegraphics[width=0.75\textwidth]{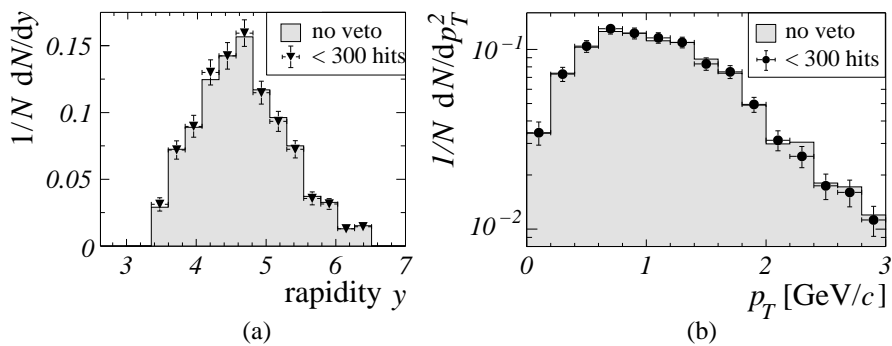}
  \caption{\emph{(a) \jpsi{} rapidity distribution for all events (histogram) 
    vs.\ events with less than 300 hits in the area covered by the 
    \rmv system (triangles). (b) \jpsi{} transverse momentum
    distribution for all events (histogram) 
    vs.\ events with less than 300 hits (circles).}}
  \label{fig:kinematics}
\end{figure}

The \jpsi{} kinematics can be described by two independent parameters. In this
analysis, the rapidity $y$ and the transverse momentum $p_\mathrm{T}$ are
chosen. The di-muon invariant mass spectrum is fitted in bins of $y$ and
$p_\mathrm{T}$, and the result using only events with less than 300 hits 
is compared  with the result using all events. The distributions are not 
corrected for detector acceptance and efficiency.
Fig.~\ref{fig:kinematics} shows the resulting rapidity and
transverse momentum distributions. Within
errors, no significant bias on the \jpsi{} kinematics is observed.

\begin{figure}[t]
  \centering
  \includegraphics[width=0.75\textwidth]{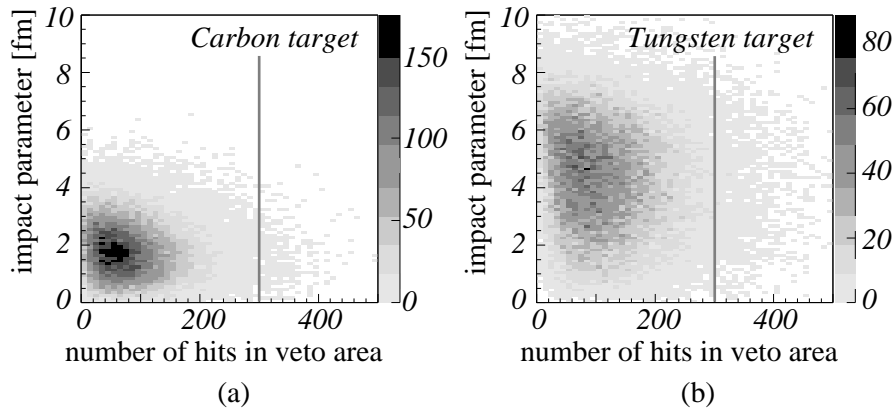}
  \caption{\emph{Expected number of hits in the area of the RICH covered by the \rmv 
    as a function of the impact parameter of the proton-nucleus
    collision for a Carbon target (a) and a Tungsten target (b).
    The data points are obtained from a Monte Carlo simulation of the
    process $\jpsi\rightarrow\mu^+\mu^-$, mixed with inelastic interactions
    according to a Poisson distribution with a mean value of 1.
    The vertical bars indicate the cut of 300 hits.}}
  \label{fig:centrality}
\end{figure}

The influence of the \rmv on the centrality of the proton-nucleus
collision in accepted events
is studied using a Monte Carlo simulation of 50,000 events each
for Carbon and Tungsten targets. The simulation
includes the process $\jpsi\rightarrow\mu^+\mu^-$ mixed with 
Poisson-distributed inelastic background with a mean value of one interaction.
In Fig.~\ref{fig:centrality}, the number of expected hits in the area of the 
RICH covered by the \rmv is shown as a function of the impact parameter of 
the proton-nucleus collision. Only a small fraction of the events
is rejected (Carbon: $<\!1$\%, Tungsten: $4$\%), and the simulation shows
no significant centrality bias.


\section{Conclusions}
Within only 1.5 years, a programmable modular veto system for the \hb
RICH has been designed, built and successfully commissioned. The system
provides a fast sum of the number of photons seen in the RICH which is used
to reject events with large multiplicities. The system shows very low bit
error rates and an efficiency close to 100\%. While the trigger and DAQ
timing is improved using the \rmv system, more than 97\%
of the \jpsi{} events survive the veto decision, and no significant bias on the
\jpsi{} kinematics is observed.

The \rmv system is used routinely in the data-taking of
the \hb experiment.

\bibliographystyle{amsplain}

\providecommand{\bysame}{\leavevmode\hbox to3em{\hrulefill}\thinspace}
\providecommand{\MR}{\relax\ifhmode\unskip\space\fi MR }
\providecommand{\MRhref}[2]{%
  \href{http://www.ams.org/mathscinet-getitem?mr=#1}{#2}
}
\providecommand{\href}[2]{#2}

\end{document}